\documentclass[%
reprint,
amsmath,amssymb,
aps,
pra,
]{revtex4-1}
 
\usepackage{graphicx}
\usepackage{dcolumn}
\usepackage{bm}
\usepackage{natbib}
\usepackage[breaklinks=true]{hyperref}

\begin{document}

\preprint{}      

\title{Continuous Production of Rovibronic Ground State RbCs Molecules via Short-Range Photoassociation to the $b^3\Pi_{1}-c^3\Sigma^+_{1}-B^1\Pi_1$ States}

\author{Toshihiko Shimasaki$^1$} 
\altaffiliation[Present address: ]{Department of Physics, University of California Santa Barbara, California 93106 USA}
\altaffiliation[]{toshihiko.shimasaki@gmail.com} 
  
\author{Jin-Tae Kim$^2$}

\altaffiliation{kimjt@chosun.ac.kr}

\author{Yuqi Zhu$^1$}

\author{David DeMille$^1$}
\altaffiliation{david.demille@yale.edu}
\affiliation{$^1$Department of Physics, Yale University, New Haven, Connecticut 06520, USA}
\affiliation{$^2$Department of Photonic Engineering, Chosun University, Gwangju 61452, Korea}




%
\date{\today}
\begin{abstract}
We have investigated rovibronic levels of the strongly mixed $b^3\Pi_1-c^3\Sigma^+_1-B^1\Pi_1$ states [the 2(1), 3(1), and 4(1) states in Hund's case (c) notation] of $^{85}$Rb$^{133}$Cs in the energy range of 13950$-$14200 cm$^{-1}$ using short-range photoassociation (PA).
For selected PA states, vibrational and rotational branching in the $X^1\Sigma^+$ state have been investigated using resonance-enhanced multiphoton ionization and depletion spectroscopy. 
Efficient production of $^{85}$Rb$^{133}$Cs molecules in the rovibronic ground state $X^1\Sigma^+ (v=0, J=0)$, at up to $\sim$1$\times 10^4$ molecules/s, has been achieved. 
\end{abstract}   

\pacs{Valid PACS appear here}
\maketitle


\section{Introduction\protect\\}  
The production of ultracold samples of heteronuclear molecules has been of widespread interest.
A variety of new applications for ultracold heteronuclear molecules have been proposed and discussed \cite{Krems2009,DeMilleReview2009}. 
With this strong motivation, many groups across the world have continued efforts toward the production of ultracold samples of heteronuclear molecules. Following early success on KRb \cite{Ni10102008}, the production of the absolute ground state of RbCs \cite{InnsbruckRbCs2014,PhysRevLett.113.255301}, NaK \cite{PhysRevLett.114.205302}, and NaRb \cite{PhysRevLett.116.205303} have been reported. These results have been accomplished by magneto-association of molecules and stimulated Raman adiabatic passage (STIRAP). This type of approach is currently most successful in terms of minimum achievable temperature and highest phase space density. However, production of ground state molecules via short-range photoassociation (PA) has the potential to be an interesting alternative: an experimental setup for PA is relatively simple, and PA can be used to produce rovibronic ground state molecules continuously. In particular, for RbCs molecules, co-trapping Cs atoms with RbCs molecules is predicted to eject unwanted molecules in the excited states, potentially leaving a pure sample of RbCs molecules in the rovibronic ground state \cite{ERHudson2008}. 

Over the past several years, a series of studies showed that short-range PA is generally possible for bi-alkali molecules: short-range PA has been demonstrated for 
NaCs \cite{PhysRevA.84.061401}, Rb$_2$ \cite{PhysRevA.87.053404,Bellos2011}, LiCs \cite{Weidemuller2008}, RbCs \cite{Bruzewicz14,PhysRevA.91.021401,Shimasaki2016}, and LiRb \cite{PhysRevA.94.062504, PhysRevA.94.062510}. Furthermore, production of rovibronic ground-state has been verified for LiCs \cite{Weidemuller2008}, RbCs \cite{PhysRevA.91.021401,Shimasaki2016}, and LiRb \cite{PhysRevA.94.062510}.

In previous work, we investigated multiple electronic states of RbCs for potentially efficient PA pathways to produce rovibronic ground state molecules. Initially, we studied the $2^3\Pi_{0^+}$ states, where we demonstrated production of the rovibronic ground state via two-photon cascade decay \cite{Bruzewicz14,PhysRevA.91.021401}. In a more recent report \cite{Shimasaki2016}, we investigated PA to the $2^1\Pi_{1}$, $2^3\Pi_{1}$ and $3^3\Sigma^+_{1}$ states. With these PA states, we achieved rovibronic ground state molecule production rate up to 2$\times10^3$ molecules/s. 
In this study, we report short-range PA to the $b^3\Pi_1-c^3\Sigma^+_1-B^1\Pi_1$ states of $^{85}$Rb$^{133}$Cs. 
This work is motivated by a recent spectroscopic study on the $B^1\Pi_1$ state of RbCs \cite{Birzniece2013}. 
In \cite{Birzniece2013}, Birzniece {\it et al.} performed Fourier-transform spectroscopy (FTS) on high-$J$ states with a resolution of $\sim$900 MHz. 
The empirical potential obtained thereby shows a clearly visible kink, which indicates strong mixing between the $B^1\Pi_1$ state and the $b^3\Pi_1$ and $c^3\Sigma^+_1$ states.
Such singlet-triplet mixing is essential for efficient production of rovibronic ground state molecules via PA: short-range PA occurs through triplet-triplet transitions around the inner classical turning point of the triplet $a^3\Sigma^+$ potential, and the PA state decays into the singlet $X^1\Sigma^+$ potential through a singlet-singlet transition. 

We investigated deeply-bound rovibrational levels of the strongly perturbed $b^3\Pi_{0^{\pm},1}-c^3\Sigma^+_{0^{-},1}-B^1\Pi_1$ states with moderately high resolution ($\sim$10 MHz), utilizing ultracold PA spectroscopy. In particular, we obtained direct information on the low rotational levels and even resolved hyperfine structures in these states. 
By using states with mixed singlet-triplet character, molecule production with rates up to $\sim$1$\times 10^4$ molecules/s into the rovibronic ground state has been achieved.

\section{The $b^3\Pi-c^3\Sigma^+-B^1\Pi$ complex of R\lowercase{b}C\lowercase{s}\protect\\}  
\begin{figure}
\includegraphics[scale=0.5]{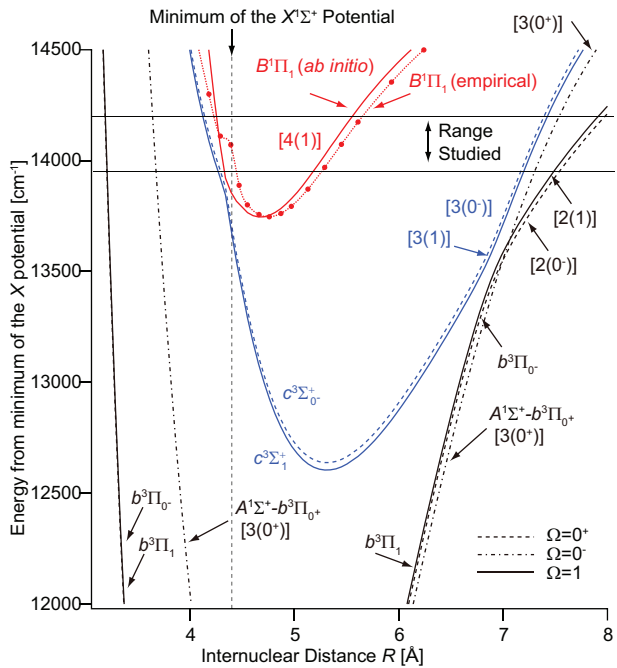}
\caption{\label{fig:PotentialZoom}(Color online) The energy region studied in this experiment. Potential curves are theoretical predictions from \cite{Allouche2002}. The empirical potential reported in \cite{Birzniece2013} is shown with dots connected by a dotted line. Hund's case (c) notation is indicated by square brackets.}
\end{figure}

We studied the energy region of $E=13950-$14200 cm$^{-1}$ above the minimum of the $X^1\Sigma^+$ potential in RbCs. 
The relevant potentials (shown in Fig. \ref{fig:PotentialZoom}) are those that are typically labelled as the $b^3\Pi_\Omega$, $c^3\Sigma^+_\Omega$, and $B^1\Pi_1$ states in Hund's case (a) notation. (High vibrational states of the $A^1\Sigma^+$ potential exist in this region as well, but do not lead to observable PA signals in this study, so we omit them here.)
The $b^3\Pi_\Omega$ and $c^3\Sigma^+_\Omega$ states split into multiple fine-structure components, namely $\Omega = 0^+,0^-,1$, and $2$ for $b^3\Pi$, and $\Omega = 0^-$ and $1$ for $c^3\Sigma^+$, respectively. Although we observed some of the $\Omega=0^{\pm}$ components as described below, our primary interest is $\Omega=1$ components: here we expect strong singlet-triplet mixing through coupling between the $b^3\Pi_{1}$, $c^3\Pi_{1}$, and $B^1\Pi_1$ states. Theoretical {\it ab initio} potential curves are available for these states \cite{Allouche2002}. 

In the past, our group performed resonance-enhanced multiphoton ionization (REMPI) spectroscopy of the $b^3\Pi_{0^\pm,1}-c^3\Sigma^+_{0^-,1}-B^1\Pi_1$ states, starting with molecules populating the $a^3\Sigma^+$ state \cite{Bergeman2004}. In that experiment, weakly-bound molecules were produced in the $a^3\Sigma^+$ potential by long-range PA, and these molecules were resonantly excited into the $b^3\Pi-c^3\Sigma^+-B^1\Pi$ complex and ionized for detection. 
However, the long-range PA populated multiple vibrational levels in the $a^3\Sigma^+$ potential that are separated by a few wavenumbers or less. Further, multiple unresolved hyperfine states of the $a^3\Sigma^+$ states contributed to the signal simultaneously. Also, the dye laser used for the $1^{\rm st}$ step of REMPI had a broad linewidth ($\sim$5 GHz), resulting in limited energy resolution. This significantly complicated the analysis of the REMPI spectra and hence information on the $b^3\Pi-c^3\Sigma^+-B^1\Pi$ energy levels. 
 
Other experimental studies for this energy region of RbCs are provided by FTS \cite{Birzniece2013, KruzinsRbCs, PhysRevA.96.022510}. In particular, in \cite{Birzniece2013}, Birzniece {\it et al.} reported a kink in the empirical potential of the $B^1\Pi_1$ state. This indicates strong mixing between the $B^1\Pi_1$ and $c^3\Sigma^+_1$ states through spin-orbit coupling. As shown in Fig. \ref{fig:PotentialZoom}, this kink was observed to occur at an internuclear distance, $R$, almost identical to that at the minimum of the $X^1\Sigma^+$ potential. Hence, vibrational levels near this kink could potentially decay efficiently to the vibrational ground state of the $X^1\Sigma^+$ potential. However, the experiment was performed with a thermal vapor, and information on the potential was obtained through extrapolation from high-$J$ data. According to Ref. \cite{Birzniece2013}, the reported frequencies for low-$J$ levels of the $B^1\Pi_1$ state have $\sim$2 cm$^{-1}$ uncertainties. More recent reports \cite{KruzinsRbCs, PhysRevA.96.022510} provided additional relevant information on the $A^1\Sigma^+-b^3\Pi$ potential curves up to $E\sim$14050 cm$^{-1}$, which has a slight overlap with the energy region investigated in this study. Some of the term values for $\Omega=0^+$ states provided in \cite{PhysRevA.96.022510} coincide with PA levels observed in our experiment.

As shown in Fig. \ref{fig:PotentialZoom}, the $b-c-B$ potentials are strongly mixed through spin-orbit interaction in the energy region above $\sim$13300 cm$^{-1}$. Therefore, we use Hund's case (c) notation [$n(\Omega)$] for PA states found in this study. States originating from the $b^3\Pi_1-c^3\Sigma^+_1-B^1\Pi_1$ complex will be labelled as 2(1), 3(1), and 4(1) states. The relevant $\Omega=0^{\pm}$ states in this energy region, which mostly originate from the $b^3\Pi_{0^-}$, $b^3\Pi_{0^+}$, and $c^3\Sigma^+_{0^-}$ states, are labelled as 2(0$^{-}$), 3(0$^{+}$), and 3(0$^{-}$) states.

\section{Experimental Scheme\protect\\}

\begin{figure}
\includegraphics[scale=0.55]{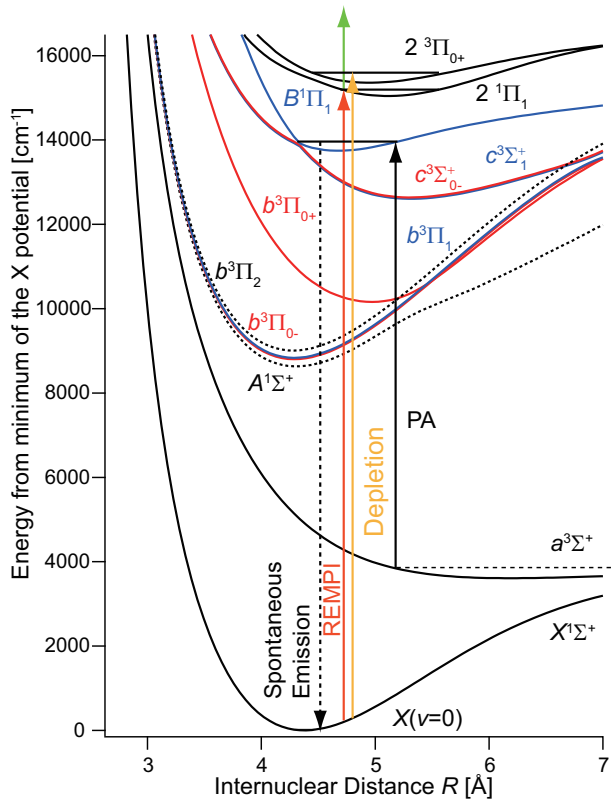}
\caption{\label{fig:scheme} (Color online) Experimental scheme for the molecule production and detection. Potential curves are from \cite{Allouche2002}. }
\end{figure}

Our experimental scheme is similar to that used in our previous studies (Fig. \ref{fig:scheme}). We start with $^{85}$Rb and $^{133}$Cs atoms co-trapped in a dual-species forced dark spontaneous-force optical trap (dark SPOT) \cite{KetterleDarkSPOT,CornellForcedDarkSPOT}.
Typically, we trap $\sim$5$\times 10^6$ Rb atoms and $\sim$1$\times 10^7$ Cs atoms with density $n_{{\rm Rb}} \sim$6$\times 10^{10}$ cm$^{-3}$ and $n_{{\rm Cs}} \sim$8$\times 10^{10}$ cm$^{-3}$, respectively, in their lowest hyperfine states.
The spatial overlap of the two atom clouds is ensured by absorption imaging from two orthogonal directions.
The translational temperature of the atoms is estimated to be $T \sim$100 $\mu$K based on time-of-flight imaging.
We irradiate the atom clouds with a PA laser, which associates pairs of colliding Rb and Cs atoms into an excited RbCs molecular state. For this PA transition, we use up to 600 mW from a diode laser, which was set up in a fiber-based master-slave system \cite{FiberInjection}.
The master laser, which consists of a 980 nm diode laser in an external cavity diode laser (ECDL) configuration, can be tuned from 10120 cm$^{-1}$ to  10350 cm$^{-1}$. The frequency of the master laser was controlled by software, using information from the transmission signal through a scanning Fabry-P\'erot cavity \cite{OrozcoCavityFeedback}. 
Typically, we were able to scan the PA laser over several GHz without an interruption such as from mode-hops.
The frequency jitter was as high as $\sim$10 MHz, which limited the frequency resolution of our PA scans. The slave laser seamlessly follows the master laser frequency by injection-locking. The PA beam is focused onto the atom clouds with a beam waist ($1/e^2$ power radius) of $\sim$150 $\mu$m. 

The excited RbCs molecules produced by PA spontaneously decay into the electronic ground state. 
We detect molecules state-selectively by using two-color (1+1$^\prime$) REMPI as shown in Fig. \ref{fig:scheme}. 
For most of this study, the wavelength of the first REMPI photon, generated from a nanosecond pulse dye laser, is tuned to 15342.0 cm$^{-1}$, which selectively excites the vibronic ground state $X(v=0)$ to the $2^1\Pi_1 (v=12)$ state \cite{Gustavsson1988}. 
The energy of the resonant pulse is kept below $\sim$0.1 mJ (with the diameter of the beam $\sim$5 mm) to reduce power broadening and off-resonance excitation.
Molecules in the intermediate excited state are ionized by an intense ($\sim$2 mJ/pulse, $\sim$5 mm in diameter) 532 nm pulse arriving 10 ns later. 
The molecular ions are detected by a Channeltron detector after acceleration by a static electric field ($\sim$100 V/cm). 
The RbCs$^+$ signals are separated from other atomic and molecular signals based on travel time to the detector.
The repetition rate of the ionization and detection is 100 Hz.
Since the linewidth of the dye laser ($\sim$5 GHz) is larger than the spacing of rotational levels, 
we use the depletion spectroscopy method to resolve rotational levels \cite{WangPRA2007, Weidemuller2008}.
The details of our depletion spectroscopy setup have been described in \cite{PhysRevA.91.021401}.

\section{\label{sec:level1}PA spectroscopy\protect\\} 
\begin{center}  
\begin{figure}[htbp]
\centering
\includegraphics[scale=0.5]{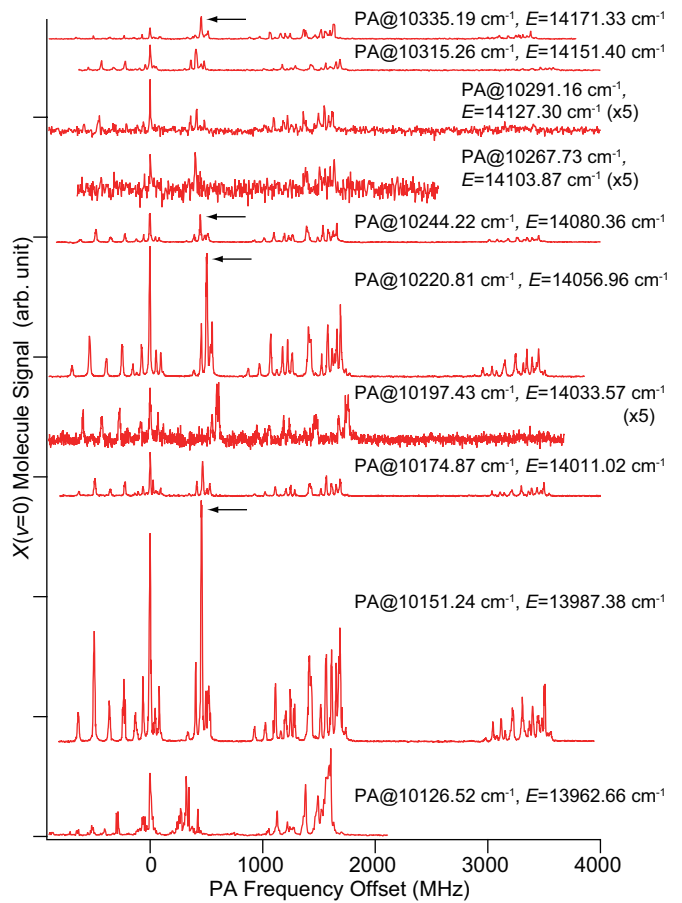}
\caption{(Color online) PA spectra of the 2(1) states in a vibrational series with spacing $\Delta E_v \simeq$ 23 cm$^{-1}$. The arrows indicate peaks used in depletion spectroscopy. The labels indicate the absolute frequency of the PA laser and the corresponding level energy at the zero of the $x$-axis for each trace. The spectra with PA at 10197.43 cm$^{-1}$, 10267.73 cm$^{-1}$, and 10291.16 cm$^{-1}$ are expanded by a factor of 5.\label{fig:PAspectra_b}} 
\end{figure}
\end{center}

\begin{center}
\begin{figure}[htbp]
\centering
\includegraphics[scale=0.5]{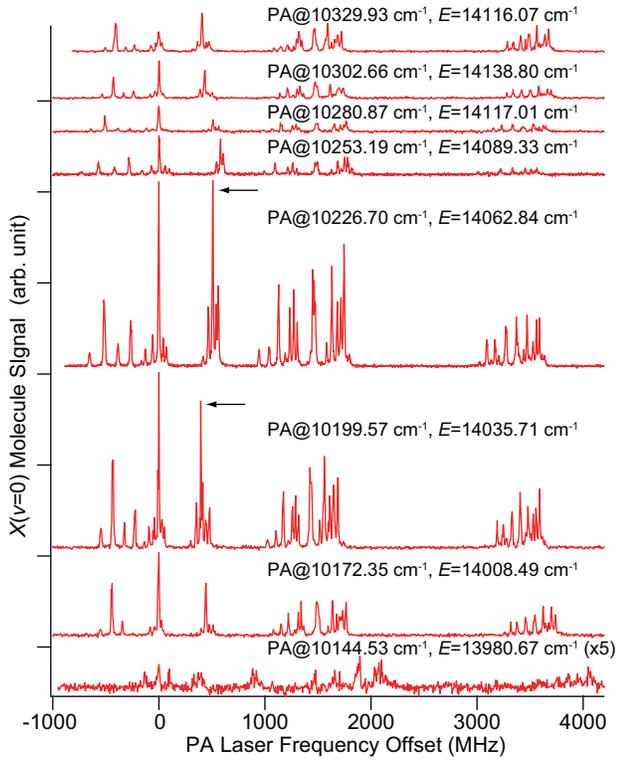}
\caption{(Color online) PA spectra of the 3(1) states in a vibrational series with spacing $\Delta E_v \simeq$ 27 cm$^{-1}$. The spectrum with PA at 10144.53 cm$^{-1}$ is expanded by a factor of 5. The arrows indicate peaks used in depletion spectroscopy. The labels indicate the absolute frequency of the PA laser and the corresponding level energy (relative to the minimum of the $X^1\Sigma^+$ state potential) at the zero of the $x$-axis for each trace. \label{fig:PAspectra_c} 
} 
\end{figure}
\end{center}

\begin{center}
\begin{figure}[htbp]
\centering
\includegraphics[scale=0.55]{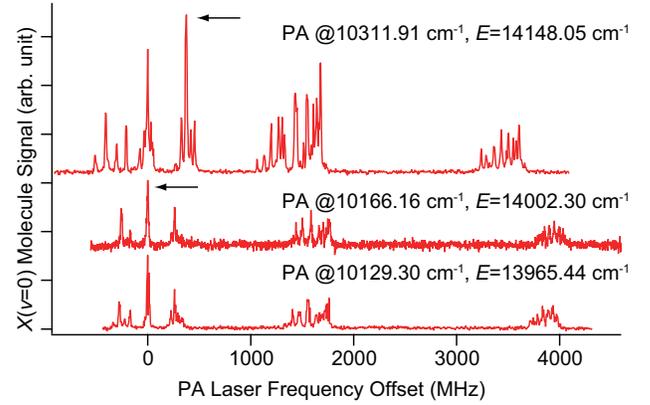}
\caption{(Color online) PA spectra of the 4(1) states ($v=$ 5, 6, and 10). The arrow indicate peaks used in depletion spectroscopy. The labels indicate the absolute frequency of the PA laser and the corresponding level energy (relative to the minimum of the $X^1\Sigma^+$ state potential) at the zero of the $x$-axis for each trace.\label{fig:PAspectra_capB}} 
\end{figure}
\end{center}

\begin{center}
\begin{figure}[htbp]   
\centering
\includegraphics[scale=0.45]{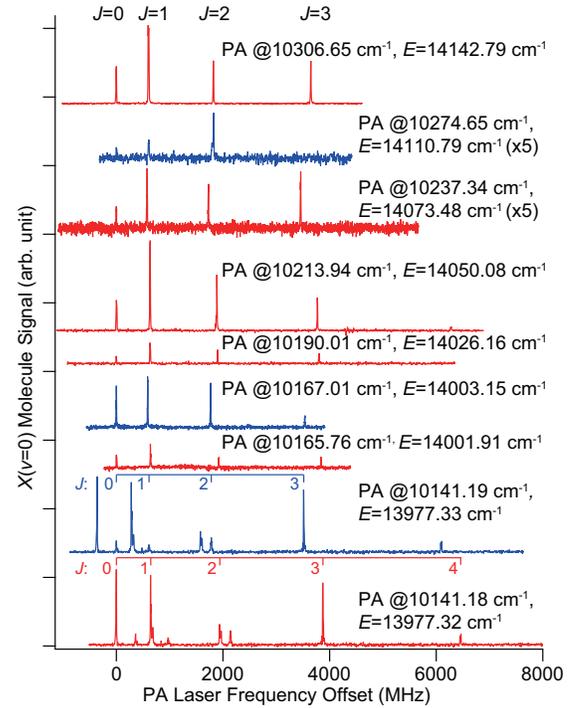}
\caption{(Color online) PA spectra of $\Omega=0^+$ and $0^-$ states. $\Omega=0^+$ and $0^-$ spectra are indicated by blue and red traces, respectively. Clear rotational progressions ($J=0,1,2$ and 3) have been observed. For the PA spectrum at 10141 cm$^{-1}$, two sets of progressions\textemdash one from an $\Omega=0^+$ state, and one from an $\Omega=0^-$ state\textemdash are observed. The labels indicate the absolute frequency of the PA laser and the corresponding level energy (relative to the minimum of the $X^1\Sigma^+$ state potential) at the zero of the $x$-axis for each trace. The spectra with PA at 10237.34 and 10274.65 cm$^{-1}$ are expanded by a factor of 5.\label{fig:PAspectra_Omega0}} 
\end{figure}
\end{center}

Due to the lack of complete knowledge about the potential curves in the energy region of interest, it is difficult to accurately predict energy level positions \textit{a priori}. 
Thus, we mostly relied on previous experiments to infer initial estimates for energy locations of the PA states. 
We first performed PA spectroscopy around the $v=8$ vibrational level of the $B^1\Pi$ state reported in \cite{Birzniece2013} to occur at $E \approx 14100$ cm$^{-1}$, where we expect a strong singlet-triplet mixing around the potential kink due to strong spin-orbit coupling. 
However, we did not observe PA lines in this energy region, possibly due to the large uncertainty in extrapolating data from the high-$J$ to low-$J$ rotational levels. 
Then we performed PA spectroscopy around the positions of the $b^3\Pi-c^3\Sigma^+-B^1\Pi$ states reported in \cite{Bergeman2004}. 
Although the frequencies reported in \cite{Bergeman2004} had some deviations from our observed line positions, we observed several PA states with hyperfine structure clearly resolved (Figs. \ref{fig:PAspectra_b}, \ref{fig:PAspectra_c}, \ref{fig:PAspectra_capB}, and \ref{fig:PAspectra_Omega0}). 

The projection quantum number $\Omega$ of the total electronic angular momentum $\bm{J_e} = \bm{L} + \bm{S}$ onto the internuclear axis can have the values of $0^+, 0^-,$ and 1 for the unbound scattering state of a pair of Rb and Cs atoms. 
According to the $ \Delta\Omega=0, \pm1$ selection rule for electric dipole transitions, we can access PA states with $\Omega=0^+, 0^-$, 1, and 2. 
For the newly observed PA lines, we identified the $\Omega$ quantum number of each PA state based on the rotational progression and hyperfine structures. 
Specifically, the existence of hyperfine structures and lack of a $J=0$ line characterizes $\Omega=1$ states. 
$\Omega=0^{\pm}$ states show a clear rotational progression without hyperfine structure, but $\Omega=0^+$ and $\Omega=0^-$ states can be distinguished by their different, characteristic patterns of relative rotational line strengths in the PA spectra \cite{Bruzewicz14,ColinThesis,Gabbanini2013}. 
The details of the spectra will be analyzed in a subsequent publication \cite{RotationalStrength}. 
Our current assignments of $\Omega$ to each PA state are given in Table \ref{tab:PAlines}, which includes updates to some of the previous assignments in \cite{Bergeman2004}. 
Our new assignments of $\Omega$ are more reliable because we resolve hyperfine structure in this study. 

\begin{center}
\begin{figure*}[htbp]
\centering
\includegraphics[scale=0.6]{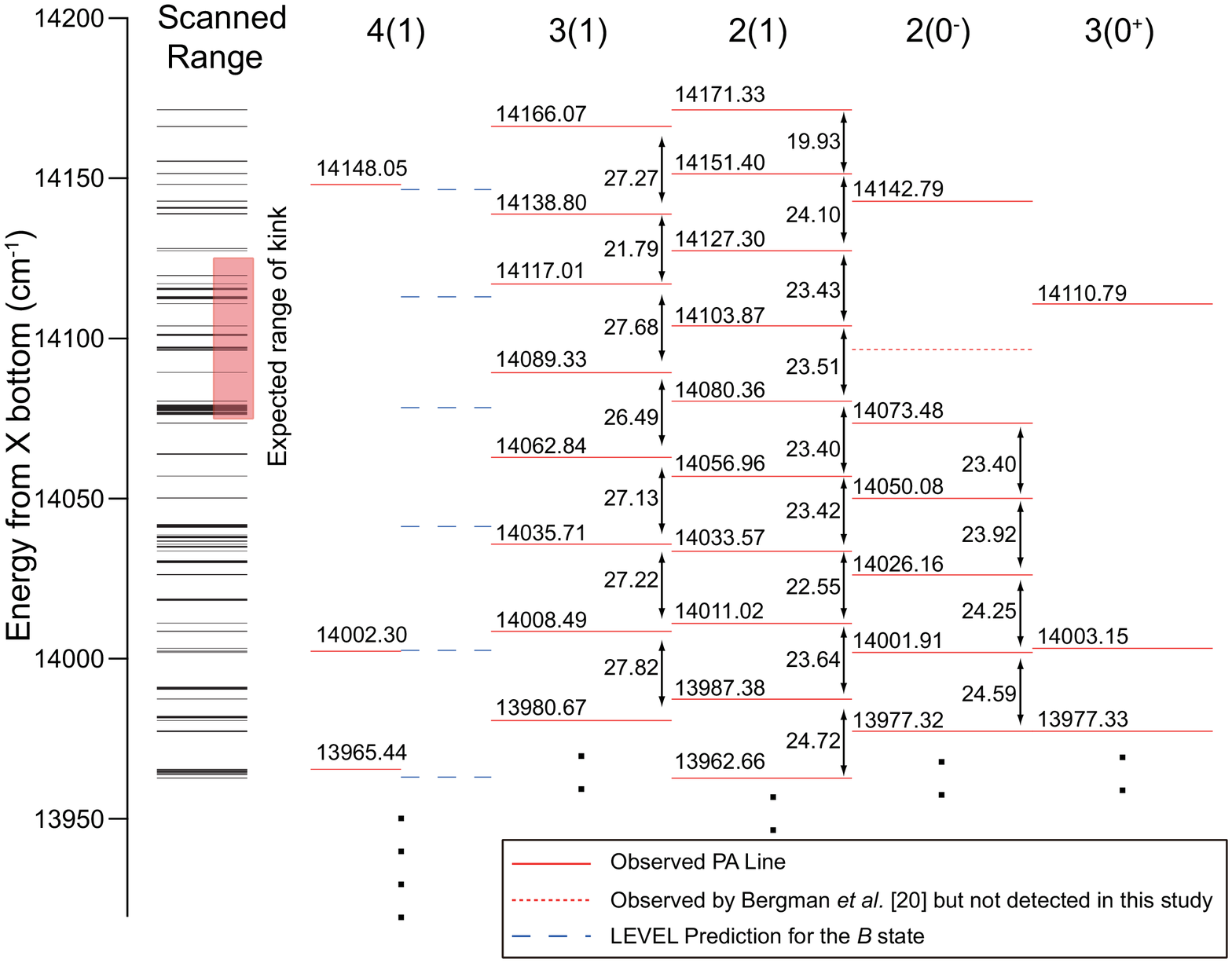}
\caption{(Color online) Summary of the observed PA lines. The states are grouped based on $\Omega$ and vibrational progressions. Energy levels and spacings between vibrational levels are given in cm$^{-1}$. Energy levels are reported as the peak at PA Laser Offset frequency equal to zero in Figs. 3-6. The energy ranges corresponding to PA scans are shown by painted rectangles. The actual frequencies for these scans are summarized in supplementary information \cite{SuppInfo}. The expected range of the kink in the empirical potential ($E\approx$14100 cm$^{-1}$) \cite{Birzniece2013} is indicated by a shaded rectangle. Several vibrational levels were reported in \cite{Bergeman2004} but were not observed in this study. One such vibrational level, which matches the vibrational progression observed here, is indicated by a dotted line. The other missing levels do not appear to match the vibrational progressions in any sensible way. These missing levels are 3-4 cm$^{-1}$ away from the levels we did observe. This is a typical value for the vibrational splittings in the populated $a^3\Sigma^+$ states that were the lower states for the transitions analyzed in \cite{Bergeman2004}. Hence, we suspect that these levels may have been assigned in error, due to a miscounting of the vibrational quantum number of the $a^3\Sigma^+$ states, in \cite{Bergeman2004}. \label{fig:EnergyDiagram} } 
\end{figure*}
\end{center}

By grouping PA states based on $\Omega$, we initially identified three sets of $\Omega=1$ series, one series of $\Omega=0^-$ states, and one set of $\Omega=0^+$ states. These series were somewhat incomplete, but allowed us to infer the location of other missing vibrational levels based on the observed progressions (Fig. \ref{fig:EnergyDiagram}). 
The PA laser was then scanned in a range around the inferred locations. With this approach, we eventually observed 30 vibronic states in this energy region in total. 
All the PA lines observed in this study are listed in Table \ref{tab:PAlines}.
However, we did not find all of the vibrational levels in these progressions due to the difficulty in scanning the narrowband ECDL over a broad frequency range. 
It should be noted that we detect only $X(v=0)$ molecules with our REMPI detection. This means that we could miss some of the PA states if they solely decay into other vibronic levels. 
\begin{table*}[htbp]
\caption{\label{tab:PAlines} Summary of the observed PA lines.}
\begin{ruledtabular}
    \begin{tabular}{cccccc}   
    Observed        & Energy from    & Normalized  &         &  Tentative State   & Rotational \\
    PA Frequency\footnote{Frequency of one of the central peaks in the lowest $J$ manifold, that is, $J=1$ for $\Omega=1$ states and $J=0$ for $\Omega=0$ states. These peaks are taken as the frequency origin in Fig. \ref{fig:PAspectra_b}, Fig. \ref{fig:PAspectra_c}, Fig. \ref{fig:PAspectra_capB} and Fig. \ref{fig:PAspectra_Omega0}. The absolute accuracy is $\sim$0.01 cm$^{-1}$.}    &  $X$ minimum\footnote{The energy of each state with respect to the minimum of the $X$ potential is obtained by adding 3836.14 cm$^{-1}$ to the PA frequency. This conversion is based on the measured value (3836.37 cm$^{-1}$) of the hyperfine-averaged dissociation energy of RbCs \cite{TiemannPhysRevA.83.052519} and the known atomic hyperfine splittings: we photoassociate a pair of $^{85}$Rb and $^{133}$Cs atoms both in their lowest hyperfine state.
}     & Signal Strength\footnote{ $X(v=0)$ molecule signal size per unit power, normalized to that for PA at 11817.15 cm$^{-1}$ [PA to the $2 ^3 \Pi_{0^+}(v=10,J=0)$ state] \cite{Bruzewicz14}, with the condition of $I \ll I_{\rm sat}$. }      &   $\Omega$       & Assignments        &  Constant\footnote{Due to the hyperfine structure, we have not extracted rotational constants from $\Omega$ =1 states.} \\
    (cm$^{-1}$) &  (cm$^{-1}$)     &       &    & in Hund's Case (c)  & (MHz) \\
    \hline
    10126.52 & 13962.66 &   4(1)   & 1     & $2(1)$ &  \\ \hline
    10129.30 & 13965.44 &   0.6(2)    & 1     & $4(1)(v=5)$  &  \\ \hline
    10141.18 & 13977.32 &   3(1)    & $0^-$    & $2(0^-)$ & 323 \\ \hline
    10141.19 & 13977.33 &   0.3(1)    & $0^+$    & $3(0^+)$ & 298 \\ \hline
    10144.53 & 13980.67 &   0.1(1)    & 1     & $3(1)$ &  \\ \hline
    10151.24 & 13987.38 &   6(2)    & 1     & $2(1)$  &  \\ \hline
    10165.76 & 14001.91 &   0.2(1)    & $0^-$    & $2(0^-)$ & 320 \\ \hline
    10166.16 & 14002.30 &   0.2(1)    & 1     & $4(1)(v=6)$  &  \\ \hline
    10167.01 & 14003.15 &   0.4(2)    & $0^+$    & $3(0^+)$ & 295 \\ \hline
    10172.35 & 14008.49 &   0.7(2)    & 1     & $3(1)$  &  \\ \hline
    10174.87 & 14011.02 &   1.0(3)    & 1     & $2(1)$  &  \\ \hline
    10190.01 & 14026.16 &   0.4(2)    & $0^-$    & $2(0^-)$ & 317 \\ \hline
    10197.43 & 14033.57 &   0.2(1)    & 1     & $2(1)$  &  \\ \hline
    10199.57 & 14035.71 &   2(1)    & 1     & $3(1)$  &  \\ \hline
    10213.94 & 14050.08 &   1.1(3)    & $0^-$    & $2(0^-)$ & 314 \\ \hline
    10220.81 & 14056.96 &   2(1)    & 1     & $2(1)$  &  \\ \hline
    10226.70 & 14062.84 &   2(1)    & 1     & $3(1)$  &  \\ \hline
    10237.34 & 14073.48 &   0.1(1)    & $0^-$    & $2(0^-)$ & 288 \\ \hline
    10244.22 & 14080.36 &   2(1)    & 1     & $2(1)$  &  \\ \hline
    10253.19 & 14089.33 &   0.2(1)    & 1     & $3(1)$  &  \\ \hline
    10267.73 & 14103.87 &   0.1(1)    & 1     & $2(1)$  &  \\ \hline
    10274.65 & 14110.79 &   0.1(1)    & $0^+$    & $3(0^+)$ & 304 \\ \hline
    10280.87 & 14117.01 &   0.5(2)    & 1     & $3(1)$  &  \\ \hline
    10291.16 & 14127.30 &   0.4(2)    & 1     & $2(1)$  &  \\ \hline
    10302.66 & 14138.80 &   0.9(3)    & 1     & $3(1)$  &  \\ \hline
    10306.65 & 14142.79 &   3(1)    & $0^-$    & $2(0^-)$ & 304 \\ \hline
    10311.91 & 14148.05 &   3(1)    & 1     & $4(1)(v=10)$  &  \\ \hline
    10315.26 & 14151.40 &   1.0(3)    & 1     & $2(1)$  &  \\ \hline
    10329.93 & 14166.07 &   0.5(2)    & 1     & $3(1)$  &  \\ \hline
    10335.19 & 14171.33 &   1.4(3)    & 1     & $2(1)$  &  \\
    \end{tabular}%
\end{ruledtabular}
\end{table*}  

Next, we attempted to assign electronic states to these series of vibrational levels. The $\Omega=0^+$ series should belong to the only $\Omega=0^+$ state, that is, the $3(0^+)$ state. Out of the three $\Omega=1$ series, the one with only three levels (shown in Fig. \ref{fig:PAspectra_capB}) is assigned to the $4(1)$ state. This is based on a comparison between our observation and the potential curve data from \cite{Birzniece2013}; we used LEVEL \cite{LeRoyLevel} with the emprical potential reported in \cite{Birzniece2013} to obtain energy levels (included in Fig. \ref{fig:EnergyDiagram}). 
The three observed levels match the levels calculated with errors up to 5 cm$^{-1}$, with a consistent vibrational spacing of $\sim$35 cm$^{-1}$. In contrast, the other two $\Omega=1$ series have much smaller vibrational spacings ($\sim$23 cm$^{-1}$ and $\sim$27 cm$^{-1}$). We similarly performed LEVEL calculations with {\it ab initio} potentials from \cite{Allouche2002} for the $2(1)$ and $3(1)$ potentials, yielding predicted vibrational spacings of $\sim$24 cm$^{-1}$ and $\sim$26 cm$^{-1}$, respectively, in the energy region studied. By comparing the vibrational spacings, the observed series with $\sim$27 cm$^{-1}$ spacing is assigned to the $3(1)$ state, and the series with $\sim$23 cm$^{-1}$ spacing should be the $2(1)$ state. Finally, by comapring the vibrational spacing of the $\Omega=0^-$ series ($\sim$24 cm$^{-1}$) with these spacings in $\Omega=1$ series, we conclude that the $\Omega=0^-$ series observed here is likely to be the $2(0^-)$ state. However, we are still missing many vibrational levels and the other $\Omega=0^-$ series (tentatively the $3(0^-)$ state); we would need more complete data sets to confirm our state assignments. 

As already mentioned, we observed clear hyperfine structures for $\Omega=1$ PA states. Interestingly, different electronic and vibrational states of the  $2(1)$, $3(1)$, and $4(1)$ states show similar hyperfine structures.
This indicates that there are likely quite strong mixings between the electronic states, notably different from the situation observed in the $2^3\Pi_{1}-3^3\Sigma^+_{1}-2^1\Pi_1$ states \cite{Shimasaki2016}. 
Analysis of these hyperfine structures will be reported elsewhere.

It is also worth mentioning that a direct decay from an $\Omega=0^-$ PA state into the $X^1\Sigma^+$ state is prohibited by the $\Omega=0^+  \not \leftrightarrow \Omega=0^-$ selection rule.
Nevertheless, we observed production of molecules in the $X$ state through $\Omega=0^-$ PA states, possibly due to a two-photon cascade decay. 
Lastly, we did not observe any $\Omega=2$ states through PA spectroscopy in this study, although PA transitions to $\Omega=2$ states are allowed and we expect the $b^3\Pi_2$ states to exist in this energy region.
Given that we observe $\Omega = 0^-$ PA states that somehow lead to population in the rovibronic ground state, there is no clear reason why $\Omega = 2$ states should not also be observable. 
Because there is no information available on $\Omega=2$ states from the previous experiment \cite{Bergeman2004}, these states may simply have fallen into gaps in our PA laser scanning range \cite{SuppInfo}. 


\section{REMPI spectroscopy\protect\\} 
\begin{center}
\begin{figure}[htbp]
\centering
\includegraphics[scale=0.45]{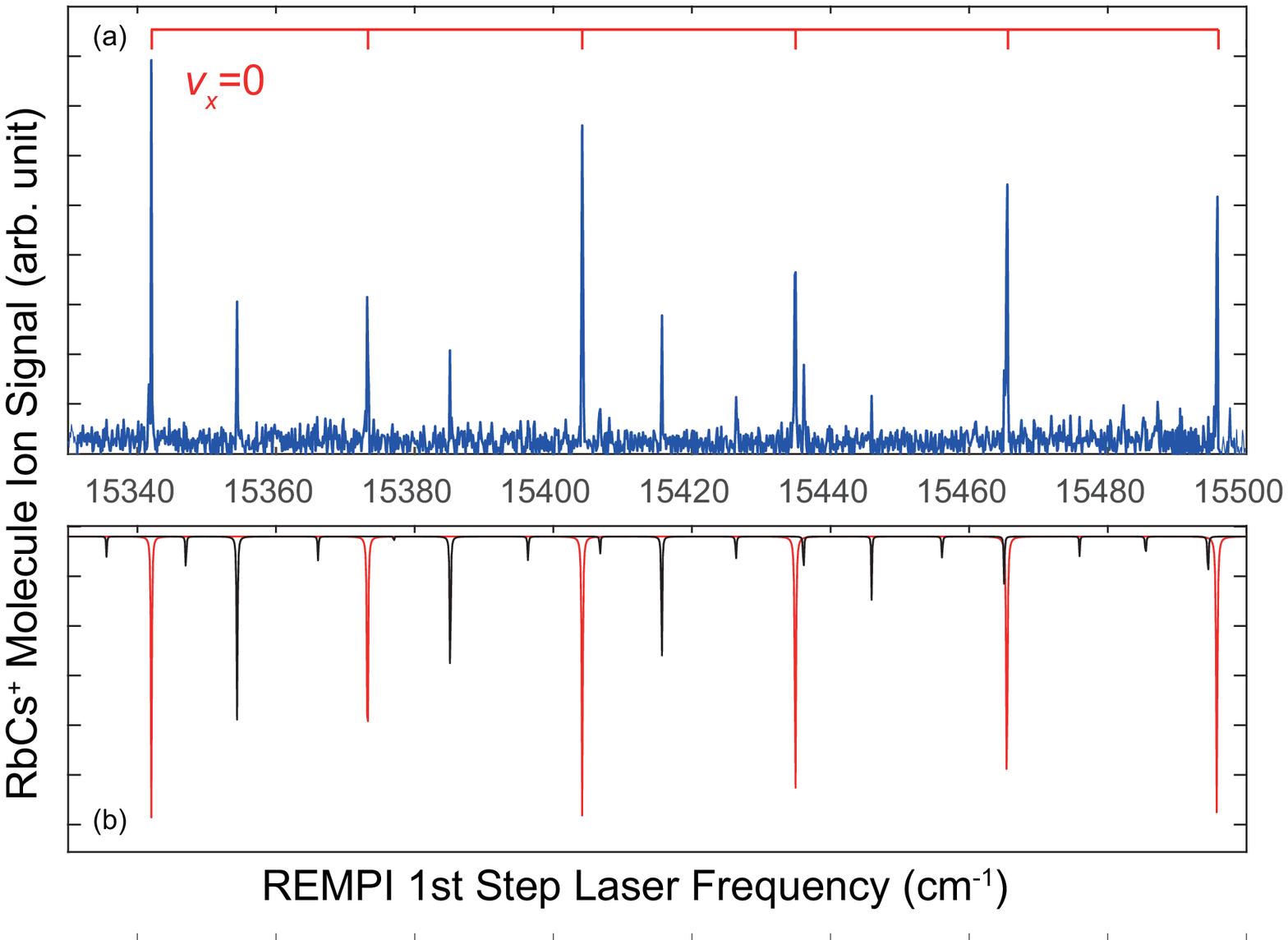}
\caption{(Color online) (a) REMPI spectrum with PA laser tuned to 10151.25 cm$^{-1}$ (the peak indicated by an arrow in Fig. \ref{fig:PAspectra_b}). (b) Simulated spectra obtained based on known REMPI transition frequencies with vibrational populations as free parameters. The red and black curves correspond to signals from molecules in the $X(v=0)$ and $X(1 \leq v \leq 3)$ states, respectively, which account for the majority of the observed peaks in the range. \label{fig:REMPI}} 
\end{figure}
\end{center}

In order to study vibrational branching in the decay of the photoassociated molecules, REMPI spectroscopy can be utilized \cite{Bruzewicz14}.
We performed similar REMPI spectroscopy for selected strong PA lines we found. 
For this type of experiment, the PA laser is fixed to a particular transition and the wavelength of the 1st step of REMPI is scanned instead. A REMPI spectrum with PA laser fixed to 10151.25 cm$^{-1}$ state is shown in Fig. \ref{fig:REMPI} as an example. By following the procedure described in \cite{Bruzewicz14}, we extracted vibrational branching ratios to low $v$ levels of the $X$ state for each PA state. Results of this analysis are summarized in Table \ref{tab:REMPI}. 

\begin{table*}[htpb]
\caption{\label{tab:REMPI} Summary of vibrational branching ratios in the decay of selected PA states. }
\begin{ruledtabular}    
\begin{tabular}{cccccccc}
    PA        & Energy from   & Tentative     &       \multicolumn{5}{c}{Vibrational Branching }  \\
    Frequency & $X$ minimum   & State    &       \multicolumn{5}{c}{ Normalized to $v_X=0$ population}  \\
    (cm$^{-1}$) & (cm$^{-1}$) &  Assignment & $v_X=1$ & $v_X=2$ & $v_X=3$ & $v_X=4$  & $v_X=5$ \\
    \hline
    10126.52 & 13962.66  & $2(1)$  & 0.2(1) & 0.2(1) & 0.2(1) & 0.1(1)  & 0.2(1) \\
    10151.24 & 13987.38  & $2(1)$  & 0.6(1) & 0.1(1) & 0.2(1) & 0.1(1)  & 0.1(1) \\
    10166.16 & 14002.30  & $4(1)$     & 0.1(2)  & 0.4(2) & 0.3(2) & 0.2(2)  & 0.1(2) \\
    10199.57 & 14035.71  & $3(1)$  & 1.2(1) & 0.2(1) & 0.1(1) & 0.1(1)  & 0.0(1) \\
    10220.81 & 14056.96  & $2(1)$  & 1.5(2) & 0.7(1) & 0.1(1) & 0.1(1) & 0.1(1) \\
    10226.70 & 14062.84  & $3(1)$  & 1.1(2) & 0.6(2) & 0.3(2) & 0.2(2) &  0.2(2) \\
    10244.22 & 14080.36  & $2(1)$  & 1.5(3) & 0.7(2) & 0.1(2) & 0.1(2)  & 0.1(2) \\
    10306.65 & 14142.79  & $2(0^-)$ & 0.5(1) & 0.2(1) & 0.2(1) & 0.1(1) & 0.2(1) \\
    10311.91 & 14148.05  & $4(1)$ & 0.3(2)  & 0.5(2) & 1.6(2) & 0.3(2) & 1.0(2) \\
\end{tabular}%
\end{ruledtabular}       
\end{table*} 

\section{Depletion spectroscopy\protect\\} 
\begin{center}
\begin{figure}[htbp]
\centering
\includegraphics[scale=0.3]{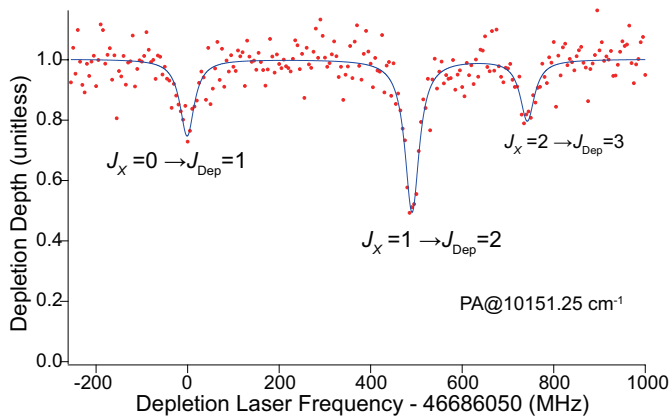}
\caption{(Color online) Depletion spectrum with PA laser tuned to 10151.25 cm$^{-1}$ (the peak indicated by an arrow in Fig. \ref{fig:PAspectra_b}). Population distributions in each rotational state in the $X(v=0)$ were obtained by curve-fitting a sum of Lorentzians (blue line) to the experimental data.\label{fig:Depletion}} 
\end{figure}
\end{center}

The relatively broad spectrum of our REMPI laser ($\sim$5 GHz) prevents us from resolving the rotational distribution of population in the vibrational ground state $X(v=0)$ simply with REMPI. As in our previous studies \cite{PhysRevA.91.021401,Shimasaki2016}, we utilize depletion spectroscopy \cite{WangPRA2007} to detect molecules in the rovibrational ground state.
The details of our depletion spectroscopy scheme can be found in \cite{PhysRevA.91.021401}. A typcial depletion spectrum, with PA laser tuned to 10151.25 cm$^{-1}$ ($J_{{\rm PA}}=1$, indicated by an arrow in Fig. \ref{fig:PAspectra_b}) is shown in Fig. \ref{fig:Depletion}. For this PA line, we observed three peaks corresponding to $J_X=0,1$ and 2 of $X(v=0)$, with a rotational branching $\sim$25\% to $J_X=0$.  
The sum of the dips is nearly 95\%, confirming the effectiveness of our molecule detection and the accuracy of the rotational branching ratios we extract. We also performed similar measurements for other strong PA lines. For $\Omega=1$ states, the central PA peaks show lower branching ratios to $J_X=0$, typically $\sim$15\%. In general, we observed $J_X$=$J_{{\rm PA}}$, $J_{{\rm PA}}\pm1$ and $J_{{\rm PA}}\pm2$. Branching to $J_{{\rm PA}}\pm2$ is usually very small, at most 10\%. We believe that $J_X=J_{{\rm PA}} \pm 2$ originates from two-photon cascade decays or mixing of different rotational levels due to hyperfine interaction. However, it is difficult to separate these two contributions with our setup.
Rotational branching ratios obtained from depletion spectroscopy data are summarized in Table \ref{tab:strongPAlines}. 

\section{Saturation behavior of new PA lines\protect\\} 
For efficient molecule production, both efficient PA and a high branching ratio to the rovibronic ground state are required. The molecule (ion) signal size (listed in Table \ref{tab:PAlines}) only gives information on the product of these two factors. Saturation intensity measurements allow us to investigate the efficiency of the PA step itself in the entire molecule production process. 

We expect the PA rate $K$ to depend on the PA laser intensity $I$ as follows \cite{PhysRevA.60.414}:
\begin{equation}\label{eq:Sat}
K(I) = 4K_{{\rm Max}} \frac{I/I_{{\rm Sat}}}{(1+I/I_{{\rm Sat}})^2}, 
\end{equation}
where $K_{{\rm Max}}$ is the maximum PA rate at the saturation intensity $I_{{\rm Sat}}$.

\begin{center}
\begin{figure}[htbp]
\centering
\includegraphics[scale=0.63]{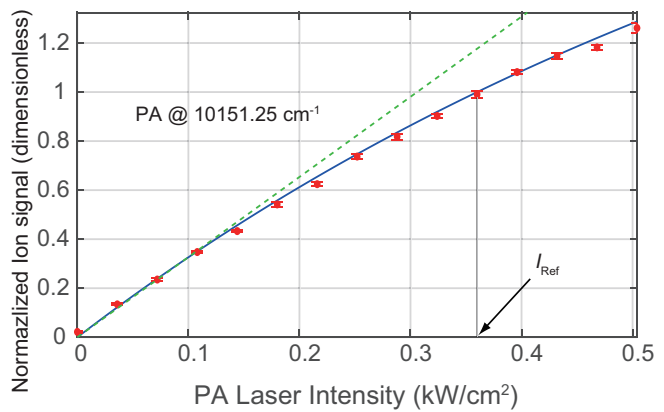}
\caption{(Color online) Saturation behavior of PA with PA laser tuned to 10151.25 cm$^{-1}$ (the peak indicated by an arrow in Fig. \ref{fig:PAspectra_b}). $I_{{\rm Ref}} = 0.36$ kW/cm$^2$, used for reference shots, is indicated by the arrow. The dotted line is a line fit to the first four data points. The curve is a fit to Eq. \ref{eq:Sat2}, with $I_{{\rm Sat}}$ as the only free parameter. $I_{{\rm Sat}} = 3 \pm 1$ kW/cm$^2$ was extracted from this data. \label{fig:Saturation}} 
\end{figure}
\end{center}

To observe saturation behavior, we performed PA with varying laser power controlled by an acoustic optic modulator (AOM). This allows us to alternate the PA intensity between the desired value and a fixed reference PA intensity, $I_{{\rm Ref}}$. By normalizing the molecule ion signal with PA intensity $I$ by another signal taken with PA intensity $I_{{\rm Ref}}$, we can minimize the effect of experimental signal size drifts. In this configuration, Eq. \ref{eq:Sat} is modified to describe the normalized PA rate $\hat{K}$, given by 
\begin{equation}
\hat{K}(I) = K(I)/K(I_{{\rm Ref}}) = \frac{I}{I_{{\rm Ref}}}\frac{(1+I_{{\rm Sat}}/I_{{\rm Ref}})^2}{(I/I_{{\rm Ref}}+I_{{\rm Sat}}/I_{{\rm Ref}})^2}. \label{eq:Sat2}
\end{equation}
Throughout these measurements, the absolute ion signal size was maintained at a low level to ensure that the ion detector response was linear even at the maximum PA power. The absolute ion signal size was reduced, if needed, by detuning the 1$^{\rm st}$-step REMPI laser a fixed distance from resonance while the PA laser power was changed.
An example PA saturation data set, taken with the PA laser tuned to 10151.25 cm$^{-1}$, is shown in Fig. \ref{fig:Saturation}. Saturation intensities, extracted by fitting the theoretical curve to experimental data, are summarized in Table \ref{tab:strongPAlines}. PA lines with higher saturation intensities are considered more promising in the long term, because it should be possible to provide a higher PA power by utilizing an enhancement cavity, and this would ultimately allow a higher molecule production rate. The maximum molecule production rates, projected based on the extracted saturation intensities, are given in Table \ref{tab:strongPAlines}.

\begin{table*}[htpb]  
\caption{\label{tab:strongPAlines} Summary of measurements on selected PA lines.}
\begin{ruledtabular}   
    \begin{tabular}{ccccccc}
                 &               &                  & Normalized       &               &   & Projected maximum \\           
    Observed     &  Energy from  & Tentative        & vibronic         &  Saturation    & Rotational & rovibronic ground-\\  
    PA Frequency &  $X$ minimum     & state            & ground state     & intensity \footnote{The uncertainties mostly originate from our limited knowledge of the PA beam size.}  & branching  & state production \\
    (cm$^{-1}$)    &  (cm$^{-1}$)     & assignment  &  production rate   & (kW/cm$^2$)   &  to $J_X=0$ \footnote{Obtained from depletion spectroscopy. For $\Omega=1$ states, the PA laser was tuned to the peaks indicated by arrows in Figs. \ref{fig:PAspectra_b}, \ref{fig:PAspectra_c}, and \ref{fig:PAspectra_capB}. For $\Omega=0^{\pm}$ states, $J=0$ PA peaks were used.  Lower bounds are given.} & rate \footnote{Estimated achievable production rate at the saturation intensity $I_{{\rm Sat}}$, for our experimental conditions.} (s$^{-1}$)\\
    \hline
    11817.15 & 15653.29 & $2^3\Pi_{0^+}$ & 1    & 5(3)      &  33 \%    & 1(1)$\times$ 10$^4$ \\
             &          & $(v=10)$ &     &      &    &  \\
    \hline
    10126.52 & 13962.66 & $2(1)$        &   4(1)     &  2(1)     &  15(5)\%  & 5(3) $\times$ 10$^3$ \\ 
    10141.18 & 13977.32 & $2(0^-)$ &   3(1)     &  10(5)    &  25(5)\%  & 2(1) $\times$ 10$^4$ \\ 
    10151.24 & 13987.38 & $3(1)$     &   6(2)     &  3(2)     &  25(5)\%  & 2(1) $\times$ 10$^4$ \\ 
    10166.16 & 14002.30 & $4(1)$        &   0.2(1)   &           &  10(5)\%  & \\ 
    10167.01 & 14003.15 & $3(0^+)$    &   0.4(2)   &           &  20(5)\%  & \\ 
    10199.57 & 14035.71 & $2(1)$        &   2(1)     &  5(3)     &  30(5)\%  & 1(1) $\times$ 10$^4$ \\ 
    10220.81 & 14056.96 & $3(1)$     &   2(1)     &  6(3)     &  25(5)\%  & 1(1) $\times$ 10$^4$ \\ 
    10226.70 & 14062.84 & $2(1)$        &   2(1)     &  4(2)     &  25(5)\%  & 7(4) $\times$ 10$^3$ \\ 
    10244.22 & 14080.36 & $3(1)$     &   2(1)     &  4(2)     &  20(5)\%  & 6(4) $\times$ 10$^3$ \\  
    10306.65 & 14142.79 & $2(0^-)$ &   3(1)     &           &  30(5)\%  & \\
    10311.91 & 14148.05 & $4(1)$        &   3(1)     &           &  20(5)\%  & \\ 
    10335.19 & 14171.33 & $3(1)$     &   1.4(3)   &           &  10(5)\%  & \\ 
    \end{tabular}%
    \end{ruledtabular}
\end{table*}  

\section{Discussion\protect\\} 

Contrary to our original expectation, the highest rovibronic ground state production was observed for triplet PA states in this energy region. 
This is probably because the short-range PA originates from the $a^3\Sigma^+$ potential. A strong coupling between the scattering state and the PA state appears to play a more important role than the coupling between the PA state and the ground state that governs the branching ratio for decay to the rovibronic ground state versus to the $a^3\Sigma^+$ state.  
Although we expect that the singlet $B^1\Pi_1$ state populates the rovibronic ground state primarily via direct (one-photon) decay from the PA state, it is difficult to determine the decay path with our experimental setup; it is still possible that the PA states studied here decay to the ground state via two-photon cascade as was observed for higher-lying PA states in \cite{PhysRevA.91.021401}.

It would be interesting to investigate other vibrational levels below the region studied here, which were not accessible to us due to tuning limitations of our PA laser. Specifically, the contribution of the $B^1\Pi_1$ state is expected to vanish below 13740 cm$^{-1}$. A study of PA in this region would help us understand the mechanism of the spontaneous decay from the simpler $b^3\Pi_1-c^3\Sigma^+_1$ complex, which then might in turn illuminate the mechanisms for decay in the full $b-c-B$ system. 

\section{Conclusions\protect\\} 
We have performed short-range PA spectroscopy of $^{85}$Rb$^{133}$Cs in the PA frequency range between 10120 cm$^{-1}$ and 10350 cm$^{-1}$ (corresponding to the energy range of $E=13950-$14200 cm$^{-1}$ above the minimum of the $X$ potential). 
30 new PA lines have been observed and assigned tentative state labels, mainly based on their vibrational progressions. 
These results supplement previous data on the energy levels of the rovibronic states of RbCs in this region. 
For selected PA states, vibrational branching and rotational branching into the $X^1\Sigma^+ (v=0)$ state have been characterized. Measurements of PA saturation behavior indicate that a higher PA power would increase the molecule production beyond what was actually observed here. 
One of the $\Omega=1$ states ($E = 13987.39$ cm$^{-1}$, PA frequency, 10151.25 cm$^{-1}$, assigned as a 2(1) state) is particularly interesting for its efficient production of the $X(v=0,J=0)$ molecules. 
The $X(v=0,J=0)$ molecule production rate we achieved for this PA state is $\sim$1$\times 10^4$ molecules/s, a factor of 5 improvement compared to previously observed production rates of rovibronic ground state RbCs molecules using PA. We project that with higher laser power, up to 2$\times 10^4$ molecules/s could be produced via PA through this state.
\begin{acknowledgments}
We thank M. A. Bellos for reading the manuscript and providing comments. This work was supported by ARO MURI. TS and JTK acknowledge support from Yale University.
\end{acknowledgments} 



\bibliographystyle{apsrev4-1} 
\bibliography{bcBPaper}

\end{document}